\documentclass[twocolumn,showpacs,aps,epsfig,nofootinbib]{revtex4}

\usepackage{graphicx}
\usepackage{epstopdf}
\usepackage{latexsym}
\usepackage{amssymb}

\usepackage[center]{subfigure}

\begin{document}

 \newcommand{\bq}{\begin{equation}}
 \newcommand{\eq}{\end{equation}}
 \newcommand{\bqn}{\begin{eqnarray}}
 \newcommand{\eqn}{\end{eqnarray}}
 \newcommand{\nb}{\nonumber}
 \newcommand{\lb}{\label}
\newcommand{\PRL}{Phys. Rev. Lett.}
\newcommand{\PL}{Phys. Lett.}
\newcommand{\PR}{Phys. Rev.}
\newcommand{\CQG}{Class. Quantum Grav.}

\title{Detailed balance condition and ultraviolet stability of   scalar field in  Horava-Lifshitz   gravity}

\author{Ahmad Borzou}
\email{ahmad_borzou@baylor.edu}

\author{Kai Lin}
\email{k_lin@baylor.edu}

\author{Anzhong Wang}
\email{anzhong_wang@baylor.edu}


\affiliation{GCAP-CASPER, Physics Department, Baylor
University, Waco, TX 76798-7316, USA}

\date{\today}

\begin{abstract}

Detailed balance and projectability conditions are two main assumptions when Horava recently formulated his theory of quantum gravity - the 
Horava-Lifshitz (HL) theory.  While the latter represents an important ingredient,   the former  often believed needs to  be abandoned, in order 
to obtain an ultraviolet stable  scalar field, among other things. In this paper, because of several attractive features of this condition, we revisit
it, and show that  the scalar field  can be stabilized, if the detailed balance  condition is allowed to be softly broken. Although this is done explicitly 
in the non-relativistic general covariant setup of Horava-Melby-Thompson  with an arbitrary coupling constant $\lambda$, generalized lately by 
da Silva, it is also true in other versions of the HL theory. With the detailed balance  condition softly breaking, the number of independent coupling 
constants can be still significantly reduced.  It is remarkable  to note that, unlike  other setups,  in this  da Silva generalization, there exists a 
master equation for  the linear perturbations of the scalar field in the flat Friedmann-Robertson-Walker  background.  
\end{abstract}


\maketitle

\section{Introduction}
\renewcommand{\theequation}{1.\arabic{equation}} \setcounter{equation}{0}

With the perspective that Lorentz symmetry may appear as an emergent symmetry at low energies, but can be fundamentally
absent at high energies,  Horava considered a gravitational system whose scaling at short
distances exhibits a strong anisotropy between space and time  \cite{Horava},
\bq
\lb{1.1}
{\bf x} \rightarrow \ell {\bf x}, \;\;\;  t \rightarrow \ell^{z} t.
\eq
In $(d+1)$-dimensions, in order for the theory to be power-counting
renormalizable,   the critical exponent $z$  needs to be $z \ge d$ \cite{Visser}.
At long distances, all the high-order curvature terms are negligible, and the linear terms become dominant. Then,
the theory is expected to  flow to the relativistic  fixed point $z = 1$, whereby the Lorentz invariance is ``accidentally restored."

The special role of time  can be  realized with the Arnowitt-Deser-Misner   decomposition  \cite{ADM},
 \bqn
 \lb{1.2}
ds^{2} &=& - N^{2}c^{2}dt^{2} + g_{ij}\left(dx^{i} + N^{i}dt\right)
     \left(dx^{j} + N^{j}dt\right), \nb\\
     & & ~~~~~~~~~~~~~~~~~~~~~~~~~~~~~~  (i, \; j = 1, 2, 3).~~~
 \eqn
 Under the rescaling (\ref{1.1})  
 with   $z = d = 3$, a condition we shall assume 
 in the rest of this paper,    $N, \; N^{i}$ and $g_{ij}$ scale,
 respectively,  as, 
 \bq
 \lb{1.3}
  N \rightarrow  N ,\;  N^{i}
\rightarrow {\ell}^{-2} N^{i},\; g_{ij} \rightarrow g_{ij}.
 \eq
 
 The gauge symmetry of the system now is reduced to the  foliation-preserving
diffeomorphisms Diff($M, \; {\cal{F}}$),
\bq
\lb{1.4}
\tilde{t} = t - f(t),\; \;\; \tilde{x}^{i}  =  {x}^{i}  - \zeta^{i}(t, {\bf x}),
\eq
under which, $N,\; N^{i}$ and $g_{ij}$ change as, 
\bqn
\lb{1.5}
\delta{g}_{ij} &=& \nabla_{i}\zeta_{j} + \nabla_{j}\zeta_{i} + f\dot{g}_{ij},\nb\\
\delta{N}_{i} &=& N_{k}\nabla_{i}\zeta^{k} + \zeta^{k}\nabla_{k}N_{i}  + g_{ik}\dot{\zeta}^{k}
+ \dot{N}_{i}f + N_{i}\dot{f}, \nb\\
\delta{N} &=& \zeta^{k}\nabla_{k}N + \dot{N}f + N\dot{f},
\eqn
where $\dot{f} \equiv df/dt$,  $\nabla_{i}$ denotes the covariant
derivative with respect to the 3-metric $g_{ij}$,   $N_{i} = g_{ik}N^{k}$, and
$\delta{g}_{ij} = \tilde{g}_{ij}(t, x^{k}) - g_{ij}(t, x^{k})$, etc.  Eq.(\ref{1.5}) shows clearly  that the
lapse function $N$ and the shift vector $N^{i}$ play the role of gauge fields of the Diff($M, \; {\cal{F}}$)
symmetry. Therefore, it is natural to assume that $N$ and $N^{i}$ inherit the same dependence on
spacetime as the corresponding generators, while    the dynamical variables $g_{ij}$ in general  
depend on both time and spatial coordinates, i.e., 
\bq
\lb{1.6}
N = N(t), \;\;\; N_{i} = N_{i}(t, x),\;\;\; g_{ij} = g_{ij}(t, x).
\eq
This is    often referred to as the {\em projectability condition}, and clearly preserved by the Diff($M, \; {\cal{F}}$).
It should be noted that, although this condition is not  necessary, breaking it often leads to 
inconsistence theories \cite{LP}.

Abandoning  the Lorentz symmetry, on the other hand,   gives rise  to   a proliferation of independently  coupling constants, 
which  could potentially limit   the prediction powers of the theory. Inspired by condensed matter systems \cite{Cardy}, 
Horava assumed that the gravitational potential ${\cal{L}}_{V}$ can be obtained from a superpotential $W_{g}$ via
the relations,
\bq
\lb{1.7}
{\cal{L}}_{V, detailed} = w^{2} E_{ij}{\cal{G}}^{ijkl}E_{kl},
\eq
where $w$ is a coupling constant, and ${\cal{G}}^{ijkl}$ denotes the generalized De Witt metric, defined as 
${\cal{G}}^{ijkl} = \big(g^{ik}g^{jl} + g^{il}g^{jk}\big)/2 - \lambda g^{ij}g^{kl}$, with $\lambda$ being another coupling constant.   
The 3-tensor $E_{ij}$ is obtained from  
 $W_{g}$ by
\bq
\lb{1.8}
E^{ij} = \frac{1}{\sqrt{g}}\frac{\delta{W}_{g}}{\delta{g}_{ij}}.
\eq
In (3+1)-dimensional spacetimes, $W_{g}$ is given by \footnote{One can include all the possible  relevant operators by adding 
$ \int{d^{3}x \sqrt{g}(R - 2\Lambda_{W})}$  to $W_{g}$. For detail, see \cite{Horava}.}, 
\bq
\lb{1.9}
W_{g}  = \int_{\Sigma}{\omega_{3}(\Gamma)},
\eq
where $\omega_{3}(\Gamma)$ denotes the gravitational Chern-Simons term,
\bq
\lb{1.10}
\omega_{3}(\Gamma) = {\mbox{Tr}} \Big(\Gamma \wedge d\Gamma + \frac{2}{3}\Gamma\wedge \Gamma\wedge \Gamma\Big).
\eq

Despite of many remarkable features  of the theory \cite{HWW,Mukc,Sotiriou,Padilla,Hreview}, it is  plagued with three major problems: 
{\em   ghost,  strong coupling and instability}. Although they are different one from another, their origins are all the same: because of
the breaking of the Lorentz symmetry, 
\bq
\lb{1.13}
\tilde{x}^{\mu} = x^{\mu}     - \zeta^{\mu}(t, {\bf x}), \; (\mu = 0, 1, 2, 3).
\eq
In particular, due to such a breaking, the kinetic part of the gravitational field in general takes the form,
\bq
\lb{1.14} 
 {\cal{L}}_{K} = K_{ij}K^{ij} -   \lambda K^{2},
 \eq
 where the extrinsic curvature $K_{ij}$ is defined as, 
 \bq
 \lb{1.15} 
K_{ij}= \frac{1}{2N}\left(- \dot{g}_{ij} + \nabla_{i}N_{j} +
\nabla_{j}N_{i}\right). 
\eq
The coupling constant $\lambda$  is subjected to radiative corrections,   and its values  are expected to  be different at different  energy 
scales. This is different from Lorentz-invariant theories, where $\lambda = 1$ is protected by the Lorentz symmetry (\ref{1.13}) even 
 in the quantum level. Then, considering linear perturbations, one can show that
the kinetic part of the gravitational  sector is proportional to $(\lambda -1)/(3\lambda - 1)$ \cite{Horava,SVW,WM,Wanga,HW}. Thus, to avoid
the ghost problem, $\lambda $ has to be either   $\lambda \ge 1$ or $ \lambda < 1/3$. 
 It is still an open question how $\lambda$ runs from its ultraviolet (UV) fixed point   to its relativistic  one, $\lambda_{IR} =1$. 
To answer this question, one way is to  study the corresponding  renormalization group (RG)  flows. However, since the  
 problem  is so much mathematically involved,  the   RG flows have not been explicitly worked out, yet, although some preliminary work has been
 already initiated  \cite{Visser,OS}. 
 
Strong coupling problem is also closely related to the fact that $\lambda$ is  generically different from one, though 
manifests itself in a different manner \cite{SC}. It appears in the self-interaction of the gravitational sector (as well as  
in the interactions of gravity with matter fields). In the framework of linear perturbations, it can be shown that  some 
coupling coefficients of third-order actions are inversely proportional to the powers of $(\lambda - 1)$ \cite{BPS,KA,PS,KP,WWb}. 
As the order increases, the  powers in terms of $1/(\lambda - 1)$ also increase \cite{Mukc}. Then, when $\lambda$ runs from its UV fixed point to its relativistic 
one, the  coupling coefficients become larger and larger. When energy is greater than the strong coupling energy scale, $\Lambda_{SC}(\lambda)$,
the coefficients become much bigger than unit, and the theory enters the strong coupling regime. 
Typically, one can show that $\Lambda_{SC} \simeq M_{pl}|\lambda - 1|^{3/4}$, where $M_{pl}$ is the Planck mass.
 Together with other observational constraints, it
implies $|\lambda - 1| \simeq 10^{-24}$  \cite{WWb}. Clearly, this gives rise to the issue of fine-tuning. 
To solve this problem, two different approaches 
have been proposed. One is the Blas-Pujolas-Sibiryakov (BPS) mechanism \cite{BPS}, in which an effective energy scale
$M_{*}$ is introduced. By properly choosing the coupling constants involved in the theory, BPS showed that $M_{*}$ can be
lower than  $\Lambda_{SC}$. As a result, the perturbative theory becomes invalid
before $\Lambda_{SC}$ is reached, whereby the problem is circumvented. While this seems a very attractive 
mechanism, it turns out  \cite{WWb}  that it may apply only to the version \cite{BPSc} of the HL theory  without projectability condition, 
a setup that also faces other challenges, including   the one with a large number ($> 60$)  of coupling constants \cite{KP}. 
The other approach is to provoke the  Vainshtein mechanism \cite{Vain}, as  showed recently in the spherical  static  \cite{Mukc} and
 cosmological \cite{WWb} spacetimes. 

Instability, on the other hand,   can appear  in  both of the  gravitational and  matter sectors. 
In the gravitational sector, due to the restricted diffeomorphisms (\ref{1.4}), 
a spin-0 graviton appears, which is unstable  in the Minkowski background \cite{Horava,SVW,WM,BS,Wang} (but, stable in the de Sitter one, as shown explicitly in \cite{WWb,Wang}),
as far as the Sotiriou, Visser and Weinfurtner (SVW) generalization with projectability condition  \cite{SVW} is concerned \footnote{In the case without projectaility 
condition, BPS showed that the instability  can be cured  by including terms made of $a_{i}$, where $ a_{i} = \partial_{i}\ln(N)$ \cite{BPSc}. However, as mentioned above,
this leads to a huge number of independent coupling constants. Then,  one may question the prediction powers of the theory. }.  This is potentially 
dangerous, and needs to decouple in the IR, in order to be consistent with observations.  It  is still an open question  whether  this is possible  or not \cite{Mukc}.

The instability of matter sector has been  mainly found for  a scalar field in the UV, due to the detailed balance condition \cite{CalA}. 
Because of this and the non-existence of relativistic  limit \cite{LMP}, it is generally believed that this condition should be abandoned 
\cite{HWW,Mukc,Sotiriou,Padilla}. 
However, the detailed balance condition has   several remarkable features \cite{Hreview}, which provoke  a great desire to revisit it.  In particular, it is in the same spirit of the 
AdS/CFT correspondence \cite{AdSCFT}, where
 a string theory and gravity defined on one space is  equivalent to a quantum field theory without gravity defined on the conformal boundary of this space, 
 which has one or more  lower  dimension(s).  Yet,
 in the non-equilibrium  thermodynamics,  the counter-part of the super potential $W_{g}$  plays the role of entropy, while the term $E^{ij}$ the entropic forces 
 \cite{OM}. This might shed light on the nature of the gravitational forces, as proposed recently  by Verlinde \cite{Verlinde}. 
 
 To overcome the above problems,   recently Horava and Melby-Thompson (HMT) \cite{HMT}   extended
the foliation-preserving-diffeomorphisms Diff($M, \; {\cal{F}}$) to include  a local $U(1)$ symmetry,
\bq
\lb{symmetry}
 U(1) \ltimes {\mbox{Diff}}(M, \; {\cal{F}}),
 \eq
 and then showed   that, similar to GR, the spin-0 graviton is eliminated \cite{HMT,WW}.
Thus, the instability of the spin-0 gravity is automatically fixed.   A remarkable by-production of this ``non-relativistic general covariant" setup is that it 
forces the coupling constant $\lambda$ to take exactly  its relativistic value $\lambda_{IR} = 1$.  Since both of the ghost and strong coupling problems
 are precisely due to the deviation of $\lambda$ from   $1$, as shown above, this in turn implies that   these two problems   are also resolved.  

However,  it was soon challenged by da Silva \cite{Silva}, who argued  that the introduction of the   Newtonian pre-potential is so strong
that actions with $\lambda \not=1$ also has the $U(1)$ symmetry. Although the spin-0 graviton is eliminated even in the da Silva generalization, as shown explicitly in 
\cite{Silva} for de Sitter and anti-de Sitter backgrounds,  and in \cite{HW} for the Minkowski, the  ghost and strong coupling problems
 rise again, since now   $\lambda$ can be different from one.   Indeed, it was shown  \cite{HW} that to avoid the ghost problem,   $\lambda$ 
must  satisfy  the same constraints, $\lambda \ge 1$ or $\lambda < 1/3$, as found previously. In addition,   the coupling becomes strong for a process
with energy   higher  than $M_{pl} |\lambda -1|^{5/4}$ in the flat Friedmann-Robertson-Walker (FRW) background, and $M_{pl}|\lambda -1|^{3/2}$  in a static weak gravitational field. 
 It must be noted that this does not contradict to the fact that the spin-0 graviton does not exist even for $\lambda \not=1$.
In fact, when one counts the degrees of freedom of the gravitational excitations, one needs to consider free gravitational fields. Another way to  count the degrees 
of the freedom is to  study the structure  of  the Hamiltonian constraints  \cite{HMT,Kluson}.
On the other hand, to study the ghost and strong coupling problems, one needs to consider  the cases in which the gravitational perturbations are different from zero. 
This can be realized by the presence of matter fields \footnote{A similar situation also 
happens in GR, in which gravitational scalar perturbations of the FRW universe in general do not vanish, although the only degrees of the freedom
of the gravitational sector  are the spin-2 massless gravitons.}. This is exactly what was done in \cite{HW}. 

It should also be noted that the HMT setup (with $\lambda = 1$) and its da Silva generalization  (with any $\lambda$) are applicable to  the cases 
with or without detailed balance condition. As a matter of fact, in \cite{HMT,Silva} the authors were mainly concerned with the theory with detailed balance condition, while
in \cite{WW,HW} the cases without detailed balance condition were studied.

Assuming that the strong coupling problem can be solved in  certain ways,  in this paper we  re-consider the stability of a scalar field with
detailed balance condition, and show explicitly that it can be stabilized in all energy scales, including the UV and IR, by softly breaking the detailed balance condition.
 We  show this explicitly in the   da Silva  generalization, although it can be easily generalized to   other versions of the HL theory. 
 It should be noted that softly breaking detailed balance condition was already considered by Horava in his seminal 
work \cite{Horava}, and later was studied by  many others, mainly in the versions without projectability condition in order to find static solutions that has relativistic limit \cite{KK,Cai}. 
We also note that    static spacetimes  were studied recently in the HMT setup (with $\lambda = 1$) \cite{AP,GSW}. 
 
The rest of the paper is organized as follows: In Sec. II  we  briefly review the da Silva generalization, and present the general action of the gravitational sector
by softly breaking the detailed balance condition. With such a breaking, the number of independent coupling constants can be still significantly reduced.  In fact, 
only in the gravitational sector, the number is already reduced from eleven to seven [cf. Eq.(\ref{2.5a})].  In Sec. III we 
study the coupling of gravity with a scalar field, and construct its most general action  
 with the same requirement:  softly breaking detailed balance condition. 
In Sec. IV, we investigate the stability of the  scalar field   with our new action of the scalar field constructed in the last section, 
and show explicitly that it is indeed stabilized in all the energy scales. 
  It is remarkable to note, unlike  other versions of the 
HL theory \cite{WWM},  now there exists a master equation for  the linear perturbations of a scalar field in the flat  FRW background for any given 
$\lambda$.  Our main conclusions are
presented in Sec. V. There are also three appendices, A, B and C. In Appendix A,    the field equations are given,  
and in Appendix
B, the linear scalar perturbations of the flat FRW universe are presented for any matter fields, while in Appendix C, the scalar perturbations for $\lambda = 1$ and $c_{1}
\not= 1$ are studied.

\section{Non-relativisitc general covariant theory with any  $\lambda$}

\renewcommand{\theequation}{2.\arabic{equation}} \setcounter{equation}{0}

In   order to limit the spin-0 graviton, HMT   introduced two new fields, the $U(1)$ gauge field $A$
and the Newtonian pre-potential $\varphi$ \footnote{Note that  the notations used  
 in this paper are   slightly different to those adopted in \cite{HMT}. In particular, we
have $\varphi = - \nu^{HMT},\; K_{ij} = - K_{ij}^{HMT},\; {\cal{A}} = - a^{HMT},\Lambda_{g} = \Omega^{HMT},\;
{\cal{G}}_{ij} = \Theta_{ij}^{HMT}$, where quantities with super indice ``HMT" are the ones used in \cite{HMT}.}.  
Under  Diff($M, \; {\cal{F}}$),
these fields transfer as,
\bqn
\lb{2.2}
\delta{A} &=& \zeta^{i}\partial_{i}A + \dot{f}A  + f\dot{A},\nb\\
\delta \varphi &=&  f \dot{\varphi} + \zeta^{i}\partial_{i}\varphi,
\eqn
while under  $U(1)$, characterized by the generator $\alpha$, 
they, together with $N,\; N^{i}$ and $g_{ij}$,
 transfer as
\bqn
\lb{2.3}
\delta_{\alpha}A &=&\dot{\alpha} - N^{i}\nabla_{i}\alpha,\;\;\;
\delta_{\alpha}\varphi = - \alpha,\nb\\
\delta_{\alpha}N_{i} &=& N\nabla_{i}\alpha,\;\;\;
\delta_{\alpha}g_{ij} = 0 = \delta_{\alpha}{N}.
\eqn
For the detail, we refer readers to \cite{HMT,WW,HW}. 

As mentioned above,   HMT considered only the case  $\lambda = 1$. Later, da Silva
generalized it to the cases with any $\lambda$ \cite{Silva}, in which the total action  can be written in the form  \cite{Silva,HW},
 \bqn \lb{2.4}
S &=& \zeta^2\int dt d^{3}x N \sqrt{g} \Big({\cal{L}}_{K} -
{\cal{L}}_{{V}} +  {\cal{L}}_{{\varphi}} +  {\cal{L}}_{{A}} +  {\cal{L}}_{{\lambda}} \nb\\
& & ~~~~~~~~~~~~~~~~~~~~~~ \left. +{\zeta^{-2}} {\cal{L}}_{M} \right),
 \eqn
where $g={\rm det}\,g_{ij}$, and
 \bqn \lb{2.5}
{\cal{L}}_{\varphi} &=&\varphi {\cal{G}}^{ij} \Big(2K_{ij} + \nabla_{i}\nabla_{j}\varphi\Big),\nb\\
{\cal{L}}_{A} &=&\frac{A}{N}\Big(2\Lambda_{g} - R\Big),\nb\\
{\cal{L}}_{\lambda} &=& \big(1-\lambda\big)\Big[\big(\Delta\varphi\big)^{2} + 2 K \Delta\varphi\Big], 
 \eqn
where $\Delta \equiv \nabla^{2} = g^{ij}\nabla_{i}\nabla_{j}$, and 
$\Lambda_{g}$ is a    coupling constant. The
Ricci and Riemann terms all refer to the 3-metric $g_{ij}$, and
 \bq
  \lb{2.6}
{\cal{G}}_{ij} = R_{ij} - \frac{1}{2}g_{ij}R + \Lambda_{g} g_{ij}.
 \eq
 ${\cal{L}}_{M}$ is the
matter Lagrangian density, which in general is a functional of the metric
and gauge fields,  
${\cal{L}}_{M} =  {\cal{L}}_{M}\big(N, \; N_{i}, \; g_{ij}, \; \varphi,\; A; \; \chi\big)$,
where $\chi$ denotes collectively the matter fields. 

${\cal{L}}_{{V}}$ is an arbitrary Diff($\Sigma$)-invariant local scalar functional
built out of the spatial metric, its Riemann tensor and spatial covariant derivatives. In Horava's original
approach  \cite{Horava}, the detailed balance condition was imposed.
His basic idea   is to assume that the potential can be constructed from a low (in the present case it is three)
dimensional super-potential $W_{g}$
via the relations of Eqs.(\ref{1.7}) - (\ref{1.10}), for which 
\bq
\lb{2.6b}
{\cal{L}}_{V, detailed} = w^{2} C_{ij}C^{ij},
\eq
where  $C_{ij}$ denotes the Cotton tensor, defined by
\bq
\lb{DBCa}
C^{ij} = \epsilon^{ikl}\nabla_{k}\Big(R^{j}_{l} - \frac{1}{4}R\delta^{j}_{l}\Big),
\eq
with $ \epsilon^{ikl} \equiv {{e}}^{ikl}/\sqrt{g}$ being the generally covariant antisymmetric tensor, and $e^{123} = 1$.
Using the Bianchi identities, one can show that
\bqn
\lb{C2}
C_{ij}C^{ij} &=& \frac{1}{8} R\Delta R - R_{ij}\Delta R^{ij} + R_{ij}\nabla_{k}\nabla^{i}R^{jk} + \nabla_{k} F^{k}\nb\\
&=& \frac{1}{2}R^{3} - \frac{5}{2}RR_{ij}R^{ij} + 3 R^{i}_{j}R^{j}_{k}R^{k}_{i}  - \frac{3}{8}\left(\nabla R\right)^{2}\nb\\
& &  +
\left(\nabla_{i}R_{jk}\right) \left(\nabla^{i}R^{jk}\right) +   \nabla_{k} G^{k},
\eqn
where $F^{k}$ and $G^{k}$ are functions of $g_{ij}$ and their derivatives only.   When integrated, they  become  boundary terms and can be discarded.

One can add relevant operators of lower dimensions to ${\cal{L}}_{{V, detailed}}$. Although in general this will break
the detailed balance condition, but such a breaking is soft, in the sense that in the UV these lower dimension operators
become negligible, and only the marginal term of Eq.(\ref{2.6b}) remains, whereby the
detailed balance condition is restored. Then, including all the relevant terms, the most general potential with the detailed balance 
condition softly breaking 
 takes the  form \cite{KKb},
 \bqn \lb{2.5a}
{\cal{L}}_{{V}} &=& \zeta^{2}\gamma_{0}  + \gamma_{1} R + \frac{1}{\zeta^{2}}
\Big(\gamma_{2}R^{2} +  \gamma_{3}  R_{ij}R^{ij}\Big)\nb\\
& &  ~~~ +  \frac{\gamma_{4}}{\zeta^{3}} \epsilon^{ijk}R_{il}\nabla_{j}R^{l}_{k} +  \frac{\gamma_{5}}{\zeta^{4}} C_{ij}C^{ij}, ~~~
 \eqn
 where the coupling  constants $ \gamma_{s}\, (s=0, 1, 2,\dots 5)$  are all dimensionless, and  ${\gamma_{5}} \equiv w^{2}{\zeta^{4}}$. The  relativistic  limit in the IR
 requires
 \bq
 \lb{gamma}
  \lambda = 1,\;\;\;
  \gamma_{1} = -1,\;\;\;
 \zeta^{2} = \frac{1}{16\pi G}. 
 \eq
 The existence of the $\gamma_{4}$ term explicitly breaks the 
 parity,  which could have  important observational consequences on primordial gravitational waves \cite{TS}.
 The corresponding field equations  are given in Appendix A. 

It should be noted that, even softly breaking the detailed balance condition, the number of the independent coupling constants is still 
considerably reduced. In fact, without detail balance condition, the number is eleven  \cite{SVW,WW} (when  the parity is allowed to be
broken), while Eq.(\ref{2.5a}) shows that  now only seven  of them remain. 

\section{Coupling of Scalar Field}

\renewcommand{\theequation}{3.\arabic{equation}} \setcounter{equation}{0}

To construct the action of a scalar field $\chi$,  following what was done  for the gravitational sector in the last section, 
we  assume that:   (i) the scalar field  respects the detailed balance condition in the UV; and 
 (ii) it breaks   the condition  only softly.  With these in mind,   let us first consider  the action,
\bqn
\lb{5.1}
\hat{S}_{\chi} &=& \frac{1}{2}\int{dt d^{3}x N\sqrt{g}\Bigg[\frac{f(\lambda)}{2N^{2}}\Big(\dot{\chi} - N^{i}\nabla_{i}\chi\Big)^{2} }
\nb\\
& &
~~~~~~~~~~~~~~~ ~~~~~~~ + \left(\frac{\delta{W_{\chi}}}{\delta\chi}\right)^{2}\Bigg],
\eqn
where,
\bqn
\lb{5.2}
W_{\chi} &=& - \frac{1}{2}\int{d^{3}x \sqrt{g}\Bigg[\sigma_{3} \chi \left(\Delta\right)^{3/2}\chi}\nb\\
& & ~~~~~~~~~~~~~~~~~~~ { + \sigma_{2} \chi \Delta\chi - m \chi^{2}\Bigg]},
\eqn
$m$ and  $\sigma_{i}\; (i = 2, 3)$  are  coupling constants. The above is quite similar to what was presented in \cite{CalA}, but with
a fundamental difference:  the sign in front of  the super-potential  now is chosen positive, opposite to what was  adopted in \cite{CalA}. 
The main reason 
 is to make the  scalar field stable in the UV, while
with the ``-" sign it was already shown that it is not  \cite{CalA}. Of course, after flipping the sign, the resulted action might not have    the desired IR limit  \cite{CalB}. 
In fact, inserting Eq.(\ref{5.2}) into Eq.(\ref{5.3}) we obtain,
\bqn
\lb{5.3}
\hat{S}_{\chi} &=& \frac{1}{2}\int{dt d^{3}x N\sqrt{g}\Bigg[\frac{f}{2N^{2}}\Big(\dot{\chi} - N^{i}\nabla_{i}\chi\Big)^{2}}\nb\\
& & ~~~~~~~~ +  \sum_{A = 2}^{6}{\beta_{A} {{P}}_{A}} + m^{2}\chi^{2}\Bigg],
\eqn
where
\bqn
\lb{5.4} 
{{P}}_{A} &=&  \chi \Delta^{A/2}\chi,\;\;\; \beta_{6} = \sigma^{2}_{3},\;\;\; \beta_{5} = 2\sigma_{2}\sigma_{3}, \nb\\
\beta_{4} &=& \sigma^{2}_{2},\;\;\; \beta_{3} = - 2m\sigma_{3},\;\;\; \beta_{2} = - 2m\sigma_{2}.
\eqn
Clearly, the mass term now has a wrong sign. Thanks to the softly breaking condition, this can be fixed by adding all the relevant terms into Eq.(\ref{5.3}).
After doing so, we find that  the scalar field can be  written  in the form,
\bq
\lb{5.5}
S_{\chi}^{(0)}\Big(N, N^{i}, g_{ij}, \chi\Big) = \int{dt d^{3}x N\sqrt{g}{\cal{L}}^{(0)}_{\chi} }, 
\eq
where
\bqn
\lb{5.6}
{\cal{L}}^{(0)}_{\chi} &=& \frac{f}{2N^{2}}\Big(\dot{\chi} - N^{i}\nabla_{i}\chi\Big)^{2} - {\cal{V}},\nb\\
{\cal V}  &=& V\left(\chi\right) + \left({1\over2}+V_{1}\left(\chi\right)\right) (\nabla\chi)^2
+  V_{2}\left(\chi\right){\cal{P}}_{1}^{2}\nb\\
& & +  V_{3}\left(\chi\right){\cal{P}}_{1}^{3}  +
V_{4}\left(\chi\right){\cal{P}}_{2} + V_{5}\left(\chi\right)(\nabla\chi)^2{\cal{P}}_{2}\nb\\
& & + V_{6} {{\cal P}}_{1} {\cal{P}}_{2},
\eqn
with $V(\chi)$ and $V_{n}(\chi)$ being arbitrary functions of $\chi$, and
\bq
\lb{5.7}
{{\cal P}}_{n} \equiv    \Delta^{n}\chi, \;\;\;
V_{6} \equiv - \sigma^{2}_{3}.
\eq

It should be noted that  the $\lambda$-dependent factor $(3\lambda - 1)$ adopted in \cite{CalA} is replaced by a  function, $f(\lambda)$,
which, subjected to some physical restrictions, is otherwise arbitrary. Those constraints include that 
the scalar field must be ghost-free in all the energy scales. 
By properly choosing it, we also assume that the speed of the scalar field  should be reduced to
the relativistic one at the IR fixed point.  

From  Eqs.(\ref{5.5})-(\ref{5.7})   one can see that the scalar field couples   directly only to   the metric components,  $N, \; N^{i}$ and $g_{ij}$.
To have it  also coupled with the gauge fields $A$ and $\varphi$, we borrow the recipe of \cite{Silva} by the repalcement, 
\bq
\lb{5.8}
S_{\chi}^{(0)} \Big(N, N^{i}, g_{ij}, \chi\Big) \rightarrow S_{\chi}\Big(N, N^{i}, g_{ij},  A, \varphi;  \; \chi\Big),
\eq
where
\bq
\lb{5.9}
 S_{\chi}  =  S_{\chi, A}(\chi, A) + S_{\chi}^{(0)} \Big(N, \big(N^{i} + N\nabla^{i}\varphi\big), g_{ij}, \chi\Big),
 \eq
and
 \bqn
 \lb{5.10}
 & & S_{\chi, A} =  \int{dt d^{3}x \sqrt{g}\Big[c_{1}(\chi)\Delta\chi + c_{2}(\chi)\big(\nabla^{i}\chi\big) \big(\nabla_{i}\chi\big)\Big]}\nb\\
 & & ~~~~~~~~~~~~~~~~~~ \times {\big(A - {\cal{A}}\big)},\nb\\
& & {\cal{A}} \equiv  - \dot{\varphi} + N^{i}\nabla_{i}\varphi + \frac{1}{2}N \big(\nabla_{i}\varphi\big)  \big(\nabla^{i}\varphi\big).
\eqn
Therefore,  the total action of the scalar field in the HMT setup can be  finally written in the form,
\bq
\lb{5.11}
S_{\chi}  =  \int{dt d^{3}x N\sqrt{g}{\cal{L}}_{\chi}}, 
\eq
where
\bqn
\lb{5.12}
{\cal{L}}_{\chi} &=& {\cal{L}}^{(0)}_{\chi} + {\cal{L}}^{(A,\varphi)}_{\chi},\nb\\
{\cal{L}}^{(A,\varphi)}_{\chi} &=& \frac{A - {\cal{A}}}{N}  \Big[c_{1}\Delta\chi + c_{2}\big(\nabla\chi\big)^{2}\Big]\nb\\
&&  - \frac{f}{N}\Big(\dot{\chi} - N^{i}\nabla_{i}\chi\Big)\big(\nabla^{k}\varphi\big)\big( \nabla_{k}\chi\big)\nb\\
& & + \frac{f}{2}\Big[\big(\nabla^{k}\varphi\big)\big(\nabla_{k}\chi\big)\Big]^{2},
 \eqn
with $ {\cal{L}}^{(0)}_{\chi}$ being given by Eq.(\ref{5.6}).

Variation of $S_{\chi}$ with respect to $\chi$ yields the generalized Klein-Gordon equation,
  \bqn
   \lb{5.16}
& & \frac{f}{N\sqrt{g}} \Bigg\{\frac{\sqrt{g}}{N}\Big[\dot{
\chi} - \left(N^{k}+N \nabla^{k} \varphi\right)\nabla_{k}\chi\Big]\Bigg\}_{,t}  \nb\\
&&  = \frac{f}{N^{2}} \nabla_{i}\Bigg\{\Big[\dot{ \chi} -\left( N^{k}+ N \nabla^{k}\varphi \right) \nabla_{k} \chi\Big] \left(N^{i}+ N \nabla^{i}\varphi \right)\Bigg\}\nb\\
 && ~~~  + \frac{g^{ij}}{N} \nabla_{i}\Bigg\{\nabla_{j}\Big[(A-{\cal{A}})c_{1}\Big]-2(A-{\cal{A}})c_{2} \nabla_{j}\chi\Bigg\} \nb\\ 
 & & ~~~ + \frac{A-{\cal{A}}}{N} \Big[c_{1}'\Delta\chi + c_{2}'\big(\nabla\chi\big)^{2}\Big]\nb\\
 & & ~~~ +  \nabla^{i}\Big[\big( 1+2V_1+2V_5 {\cal P}_2 \big)\nabla_{i} \chi\Big] \nb\\
 && ~~~  -  {\cal V}_{, \chi} - \Delta\left({\cal V}_{,1}\right) - \Delta^{2}\left({\cal V}_{,2}\right),
  \eqn
where $c_{1}' \equiv dc_{1}(\chi)/d\chi$, etc, and
  \bqn
 \lb{5.17}
{\cal V}_{, \chi} &\equiv& \frac{\partial {\cal V}}{\partial \chi} =  V'  + V_{1}' (\nabla\chi)^2
   +  V_{2}'{\cal{P}}_{1}^{2}  +  V_{3}'{\cal{P}}_{1}^{3} \nb\\
  & &  ~~~~~~~~~   + V_{4}'{\cal{P}}_{2} + V_{5}'
  (\nabla\chi)^2{\cal{P}}_{2}. 
 \eqn

\section{Stability of Scalar Field}

\renewcommand{\theequation}{4.\arabic{equation}} \setcounter{equation}{0}

In this section, we consider the problem of stability of the scalar field in a flat FRW background,
\bq
\lb{3.1}
\hat{N} = a(\eta),\;\; \hat{N}_{i} = 0,\;\; 
\hat{g}_{ij} = a^{2}(\eta)\delta_{ij}.
\eq
Since in this paper  we are working with the conformal time $\eta$, we use symbols with hats to denote the 
quantities of background, in order to distinguish from the ones used in the coordinates ($t, x^{i}$)  \cite{WM,WW,HW}, 
where $t$ denotes    the cosmic time.  The relations between the two difference coordinate systems are given 
explicitly in \cite{WW,HW}.  

The flat FRW universe and its linear perturbations with any matter fields are presented in  Appendix B. In this section, as well as 
in the next, we shall  apply those formulas to the case where the only source is a scalar field, constructed in the last section.

In particular, without loss of generality,
we assume that
$\hat{A} = \hat{A}(\eta), \; \hat{\varphi} = \hat{\varphi}(\eta)$.
However, using the $U(1)$ gauge freedom of Eq.(\ref{2.3}), we can always set one of them zero.  In this paper,
  we choose the gauge
\bq
\lb{gauge}
\hat{\varphi} = 0.
\eq

Thus, to zero-order, from Eq.(\ref{5.13a}) -(\ref{5.14}) we find that
\bqn
\lb{equ1}
 \hat{J}^{t} &=& - 2\Bigg(\frac{f}{2a^{2}} \hat{\chi}'^{2} + V(\hat{\chi})\Bigg), \nb\\
\hat{J}^{i} &=& 0, \;\;\; \hat{J}_{\varphi}  = \hat{J}_{A}  = 0, \nb\\
 \hat{\tau}_{ij} &=&\Bigg(\frac{f}{2a^{2}} \hat{\chi}'^{2} - V(\hat{\chi})\Bigg)a^{2} \delta_{ij},
  \eqn
where a prime denotes the ordinary derivative with respect to its indicated argument, for example,
$ \hat{\chi}' \equiv d \hat{\chi}/d\eta,\; \hat{V}' \equiv d\hat{V}/d\hat{\chi}$, etc. 
Hence, the generalized Friedmann equation (\ref{3.7a}) and the conservation law of energy (\ref{3.8}) become, respectively,
\bqn
\lb{equ2}
& &   \frac{{\cal{H}}^{2}}{a^{2}}   = \frac{8\pi \hat{G}}{3}\Bigg(\frac{1}{2a^{2}} \hat{\chi}'^{2} + \hat{V}(\hat{\chi})\Bigg),\\
\lb{equ3}
& & \hat{\chi}'' + 2{\cal{H}}\hat{\chi}' + a^{2} \hat{V}'(\hat{\chi}) = 0, 
\eqn
where 
\bq
\lb{equ3a}
\hat{G} \equiv \frac{2fG}{3\lambda - 1},\;\;\;
\hat{V} \equiv \frac{V}{f}.
\eq
It should be noted that in writing Eqs.(\ref{equ2}) and (\ref{equ3}), we had set $\Lambda = 0$, where $\Lambda \equiv \gamma_{0}\zeta^{2}/2$. From Eqs.(\ref{3.8a}) and (\ref{3.8b}), 
on the other hand, we also obtain $\Lambda_{g} = 0$. 
 Note that   Eq.(\ref{equ3}) can be  obtained from the generalized Klein-Gordon equation (\ref{5.16}). In addition, the gauge field
 $A$ is not determined.  One may set $\hat{A} = 0$. However, in this paper we shall leave this possibility 
 open. 
 
 It is remarkable that Eqs.(\ref{equ2}) and (\ref{equ3})
are the same as those given in GR, after    replacing    $(\tilde{G}, \tilde{V})$ by $ (G, V)$. 
Thus, all the results obtained there are equally applicable to  the present case, as
far as only the background is concerned.  These include the slow-roll conditions for inflation \cite{MW09},
\bq \lb{equ3b}
\epsilon_{V},\;
 |\eta_{V}| \ll 1,
\eq
where
 \bq \lb{equ3c}
\epsilon_{V} \equiv \frac{1}{16\pi \tilde{G}}\, \frac{\tilde{V}^{\prime 2}
}{\tilde{V}^2},\;\;\; \eta_{V} \equiv \frac{1}{8\pi \tilde{G}}\,
\frac{\tilde{V}''}{\tilde{V}}. 
\eq
However, because of the presence of high-order spatial derivatives here, 
the perturbations are expected to be dramatically different, as shown below.

The linear scalar perturbations are given by
\bqn
\lb{4.0a}
\delta{N} &=& a \phi,\;\;\; \delta{N}_{i} = a^{2}B_{,i},\nb\\
\delta{g}_{ij} &=& -2a^{2}\big(\psi \delta_{ij} - E_{,ij}\big),\nb\\
A &=& \hat{A} + \delta{A},\;\;\; \varphi = \hat{\varphi} + \delta\varphi.
\eqn
Using the $U(1)$ gauge freedom (\ref{4.0c}), we can further set,
\bq
\lb{4.0dd}
{\delta\varphi} = 0.
\eq 
In addition, we also adopt the quasi-longitudinal gauge \cite{WM},
\bq
\lb{4.0de}
\phi = 0 = E.
\eq
With the above gauge choice, it can be shown that the gauge freedom is completely fixed. Then, the linear perturbations for any matter fields are given in
Appendix B. When the scalar field is the only source of the spacetime,  to first-order,  Eq.(\ref{5.13a}) -(\ref{5.14}) yield
\bqn
\lb{equ4}
 \delta{\mu} &=& 
 \frac{f}{a^{2}}\hat{\chi}'\delta\chi' + \Bigg(V' + \frac{V_{4}}{a^{4}}\partial^{4}\Bigg)\delta\chi, \nb\\
\delta{J}^{i} &=&  \frac{f}{a^{3}}\hat{\chi}'\partial^{i}\delta\chi,\nb\\
\delta{J}_{A}  &=& \frac{2}{a^{2}}c_{1}\partial^{2}\delta\chi,  \nb\\
 \delta{\tau}_{ij} &=& \Bigg[f\hat{\chi}'\Big(\delta\chi' - \hat{\chi}'\psi\Big) - a^{2}\Big(V'\delta\chi - 2V\psi\Big)\Bigg]\delta_{ij},  \nb\\
\delta{J}_{\varphi} &=& \frac{1}{a^{4}}\Big(ac_{1}\partial^{2}\delta\chi\Big)' - \frac{f}{a^{3}}\hat{\chi}' \partial^{2}\delta\chi.
 \eqn
Then, Eqs.(\ref{4.5}) and  (\ref{4.6a}) - (\ref{4.7b}) reduce, respectively, to
\bqn
\lb{d.1}
& & (3\lambda - 1)\psi'  + (\lambda - 1)\partial^{2}B =  8\pi fG \hat{\chi}' \delta\chi,\\
\lb{d.2}
& & 2{\cal{H}}\partial^{2}\psi -   (\lambda - 1)\partial^{2}(3\psi' + \partial^{2}B)  \nb\\
& & ~~~~~~~~~~ = 8\pi G\Big[\big(c_{1}' -f \hat{\chi}' + {\cal{H}}c_{1}\big)\partial^{2}\delta\chi   + c_{1}\partial^{2}\delta\chi'\Big],~~~~~~~~~~~\\
\lb{d.3}
&&  \psi  =  4\pi G c_{1}\delta\chi,\\
\lb{d.4}
&& \psi'' + 2{\cal{H}} \psi'  + \frac{\lambda -1}{3\lambda -1}\partial^{2}\big(B' + 2{\cal{H}}B\big) \nb\\
& & ~~ =  \frac{8\pi G}{3\lambda - 1}\Big(f \hat{\chi}' \delta\chi' - a^{2}V' \delta\chi\Big), \\
\lb{d.5}
& & B' + 2{\cal{H}}B  = \Bigg(\frac{a -  \hat{A}}{a} + \frac{8\gamma_{2} + 3\gamma_{3}}{\zeta^{2} a^{2}}\partial^{2}\Bigg)\psi  + \frac{\delta{A}}{a}. ~~~
\eqn
It can be shown that Eq.(\ref{d.2}) is not independent, and can be derived from the rest.  

 On the other hand, 
 the  Hamiltonian constraint Eqs.(\ref{4.4})  and the conservation law of energy  (\ref{4.9a}) become,
\bqn
\lb{d.6}
& & \int{dx^{3} \Bigg[\partial^{2}\psi -  \frac{1}{2}(3\lambda -1) {\cal{H}}\big(3\psi' + \partial^{2}B\big)\Bigg]}\nb\\
& & ~~~~~ =  4\pi G \int{dx^{3}\Bigg[f\hat{\chi}'\delta\chi'   + a^{2}V'\delta\chi  + \frac{V_{4}}{a^{2}}\partial^{4}\delta\chi\Bigg]},~~~~~~~~ \\
 \lb{d.7}
 & & \int{ d^{3}x \Bigg\{fa^{2}\Big(\hat{\chi}'' \delta\chi'  + \hat{\chi}' \delta\chi'' - 2{\cal{H}}\hat{\chi}' \delta\chi' - 3 \hat{\chi}'^{2}\psi'\Big) }\nb\\
& & ~~~ +\big(V_{4}' + 3{\cal{H}}V_{4}\big)\partial^{4}\delta\chi - \big(ac_{1} \hat{A} - V_{4}\partial^{2}\big)  \partial^{2}\delta\chi'  \nb\\
 & & ~~~ +   \Big[a^{4}\hat{\chi}'V'' - a\big(\hat{\chi}'c_{1}' + 3{\cal{H}}c_{1}\big)\hat{A}\partial^{2}\Big] \delta\chi\Bigg\} = 0.
 \eqn
The conservation law of momentum (\ref{4.9b}) is satisfied identically, and the Klein-Gordon equation (\ref{5.16}) takes the form,
\bqn
\lb{d.8}
& &f\Big[\delta\chi'' + 2{\cal{H}}\delta\chi'  - \hat{\chi}'\big(3\psi' + \partial^{2}B\big)\Big] + a^{2}V''\delta\chi  \nb\\
& & ~~~~= \frac{c_{1}}{a}\partial^{2}\delta{A} + \frac{1}{a}\Big[a\big(1 + 2V_{1}\big) + 2\hat{A}\big(c_{1}' - c_{2}\big)\Big]\partial^{2}\delta\chi\nb\\
& &~~~~~~  - \frac{2}{a^{2}} \big(V_{2} + V_{4}'\big) \partial^{4}\delta\chi + \frac{2\sigma^{2}_{3}}{a^{4}}  \partial^{6}\delta\chi.
\eqn
 To further study it, it is found convenient to  consider the cases  $\lambda = 1$ and $\lambda \not=1$ separately.   
  
\subsection{$\lambda \not=1$}

When $\lambda \not=1$, substituting Eqs.(\ref{d.1}), (\ref{d.3}) and (\ref{d.5}) into Eq.(\ref{d.8}), we find that in the moment space it can be cast in the form,
\bq
\lb{equ8}
 u_{k}'' + \left(\omega^{2}_{k}-  \frac{a''}{a}\right)u_{k} = 0,
\eq
where $\delta\chi = u/a$, and
\bqn
\lb{equ9}
  \omega^{2}_{k} &=&   \frac{a^{\prime\prime}}{a} 
   +\frac{4a_{2}a_{3} + 2a_{1}a_{3}^\prime-a_{1}^2-2a_{3}a_{1}^\prime}{4a_{3}^2},\nb\\
 a_{1} &\equiv& 2{\cal{H}} +8\pi G\frac{3\lambda-1}{f(\lambda-1)}c_1\big(\hat{\chi}^\prime c_1^\prime+c_1{\cal{H}}\big),\nb\\
a_{2} &\equiv& \frac{2\sigma^2_3k^6}{fa^4}  +\frac{2k^4}{fa^2}\left[V_2+V_4^\prime+2\pi Gc^2_1\frac{8\gamma_2+3\gamma_3}{\xi^2}\right]\nb\\
& & +\frac{k^2}{fa}\Big[a(1+2V_1)-\hat{A}(c_1^\prime-c_2)-4\pi Gc_1^2(a-\hat{A})\Big]\nb\\
& & - \frac{8\pi G}{\lambda-1}\Bigg[c_1\hat{\chi}^{\prime\prime} + \left(f-c_1^\prime\right)\left(\hat{\chi}'\right)^{2} 
-\frac{3\lambda-1}{2f}c_1c_1^\prime\hat{\chi}^{\prime}\nb\\
& & ~~~~~~~~~~~ ~ +2c_1{\cal{H}}\hat{\chi}^{\prime}\Bigg]
 +\frac{a^2V^{\prime\prime}}{f}, \nb\\
a_{3}&\equiv&1+4\pi Gc^2_1\frac{3\lambda-1}{f(\lambda-1)}.
 \eqn
 It is remarkable that, in contrast to other version of the HL theory \cite{WWM}, now a master  equation exists.   
 In the UV ($ k \gg  {\cal{H}}$),  we have $ \omega^{2}_{k} \simeq  {2\sigma^2_3k^6}/(fa^4) > 0$, and the scalar field is indeed stabilized. In the IR ($ k \ll  {\cal{H}}$),
the last term in the expression of $a_{2}$ dominates. Since $V'' = 2 m_{\chi}^{2} > 0$, where  $m_{\chi}$ denotes the mass of the scalar field, one can see that it is stable also
in the IR.

In particular,  in the extreme slow-roll (de Sitter) limit, we take $\hat{\chi}' \simeq V'(\hat{\chi}) \simeq 0$ 
and $a  \simeq -1/(H\eta)$, where $H = [8\pi \hat{G}\hat{V}(\hat{\chi}_{0})/3]^{1/2}$. Then, Eq.(\ref{equ9}) reduces to,
 \bqn
 \lb{equ10}
   \omega^{2}_{k} &=&\frac{1}{a_{3}}\Bigg\{\frac{2\sigma^2_3k^6}{fa^4}  +\frac{2k^4}{fa^2}\left[V_2+V_4^\prime+2\pi Gc^2_1\frac{8\gamma_2+3\gamma_3}{\xi^2}\right]\nb\\
& & +\frac{k^2}{fa}\Big[a(1+2V_1)-\hat{A}(c_1^\prime-c_2)-4\pi Gc_1^2(a-\hat{A})\Big]\nb\\ 
& & ~~~~~~~~~~~ ~  
 +\frac{a^2V^{\prime\prime}}{f}\Bigg\}, \; (\hat{\chi}' = 0),
  \eqn
which clearly shows that the scalar field is stable in all the energy scalar, by properly choosing the potential terms $V_{i}$.

\subsection{ $\lambda = 1$} 

In this case, if $c_{1} = 0$,   Eqs.(\ref{d.1}) and (\ref{d.3}) imply
$\hat{\chi}' = 0 = \psi$, 
 that is, {\em to zero-order the scalar field   must    take its vacuum expectation value, $\hat{\chi} = \hat{\chi}_{0}$, where $V'(\hat{\chi}_{0}) = 0$, and the 
corresponding background is necessarily  de Sitter}.
Then, Eqs.(\ref{d.5}) reads,
\bq
\lb{c.12}
\delta{A} = \frac{1}{a}\big(a^{2}B\big)'.
\eq
The Klein-Gordon equation (\ref{d.8}) can be also written in the form of Eq.(\ref{equ8}), but now with 
\bqn
\lb{c.13}
  \omega^{2}_{k} &=& \frac{V''}{fH^{2}\eta^{2}}  + \frac{1}{f}\left[\big(1 + 2V_{1}\big) + 2c_{2}H\hat{A}\eta\right]k^{2}\nb\\
& & + \frac{2H^{2}\eta^{2}}{f}\Big[\big(V_{2} + V_{4}'\big) + \sigma^{2}_{3}H^{2}\eta^{2}k^{2}\Big]k^{4}.
\eqn
Again, since $V'' = 2m_{\chi}^{2} > 0$,  the above show that the scalar field is stable in the IR, by properly choosing
the potential term $V_{1}$. In the UV,  the $\sigma^{2}_{3}$ dominates, and is strictly positive, so it is also stable in this regime. In fact, by properly 
choosing the potentials $V_{2}$ and $V_{4}$, it can be made stable in all the energy scales.  

It is interesting to note  that in this case, similar to the background,  the gauge field $A$ is not determined by the field equations. 

When $c_{1} \not= 0$,  one can find explicit solutions for $\delta \chi,\;  \psi,\; B$ and $\delta{A}$, and are given in Appendix C,
which seem not physically much interesting. So,   we shall not consider them further.

\section{Conclusions}

\renewcommand{\theequation}{6.\arabic{equation}} \setcounter{equation}{0}

In this paper, we have studied the stability of a scalar field in the case where the detailed balance condition is softly broken.
 This is done in the da Silva generalization 
\cite{Silva} of the   HMT setup \cite{HMT}.

In Sec. III, we have first constructed the most general action of a scalar field with softly breaking the detailed balance condition, while
 in Sec. IV, we have studied its stability and shown explicitly  that    the scalar field becomes stable in all the energy scales, including 
 the UV and IR.     It is remarkable to note that, unlike  other versions of the HL theory \cite{WWM}, there exists a master equation for  the 
  linear perturbations of a scalar field  in the flat FRW  universe   in this setup.
 
It  should be also noted that  our conclusions do not contradict to the ones obtained earlier by Calcagni \cite{CalA,CalB}, who showed that
a scalar field is not stable in the UV when the detailed balance condition is imposed. This is mainly due to two facts: First, we have chosen
a different sign in the front of the super-potential of the scalar field [cf. (\ref{5.1})]. This helps to improve the UV stability of the scalar field,
but usually leads to undesired behavior in the IR \cite{CalB}. To fix the latter, we allow the detailed balance condition to be softly broken,
so that the behavior of the scalar field in the IR is healthy.  Such a breaking is also desired by other considerations  \cite{LMP,KK,Cai}, 
including the solar system tests. In addition, it
still provides a very  effective mechanism to reduce significantly  the number of independent coupling constants of the marginal terms. For example, in the 
gravitational sector, this condition reduces   the number from five ($g_{4}, g_{5}, ..., g_{8}$, as given in \cite{SVW,WM})  to one, denoted by
$\gamma_{5}$ in   Eq.(\ref{2.5a}). Moreover, it may also shed light on the nature of the gravitational forces \cite{Hreview,AdSCFT,Verlinde}.

In addition, our conclusions regarding to the stability of a scalar field can be easily generalized to other versions of the 
HL theory \cite{Mukc,Sotiriou,Hreview}, including the SVW setup \cite{SVW}.

Finally, we note that there does not exist the ghost problem for the scalar field $\chi$, provided that $f(\lambda) > 0$. This can be seen from 
Eqs.(\ref{5.6}), (\ref{5.11}) and (\ref{5.12}), where the kinetic part (in the gauge $\varphi = 0$) reads
$$
{\cal{L}}_{\chi}^{(K)} = \frac{f}{2N^{2}}\left(\dot{\chi} - N^{i}\nabla_{i}\chi\right)^{2}\nb\\
\simeq \frac{f}{2a^{2}}\left(\hat{\chi}' + \delta\chi'\right)^{2},
$$
to the first-order approximations. Clearly, it is always positive for $f > 0$.

~\\{\bf Acknowledgements:}   
The authors would like to thank G. Calcagni and Y.-Q. Huang for valuable discussions and comments. 
The work of AW was supported in part by DOE  Grant, DE-FG02-10ER41692.

 \section*{Appendix A: The field Equations} 
 
 \renewcommand{\theequation}{A.\arabic{equation}} \setcounter{equation}{0}
 
 Variation of the total action (\ref{2.4}) with respect to   $N(t)$  yields the
Hamiltonian constraint,
 \bqn
 \lb{eq1}
& & \int{ d^{3}x\sqrt{g}\left[{\cal{L}}_{K} + {\cal{L}}_{{V}} - \varphi {\cal{G}}^{ij}\nabla_{i}\nabla_{j}\varphi
- \big(1-\lambda\big)\big(\Delta\varphi\big)^{2}\right]}\nb\\
& & ~~~~~~~~~~~~~~~~~~~~~~~~~~~~~
= 8\pi G \int d^{3}x {\sqrt{g}\, J^{t}},
\eqn
where
 \bq
  \lb{eq2b}
  J^{t} \equiv 2 \frac{\delta\left(N{\cal{L}}_{M}\right)}{\delta N}.
 \eq
Variation with respect to  $N^{i}$ yields the super-momentum constraint,
 \bq \lb{eq2}
 \nabla^{j}\Big[\pi_{ij} - \varphi  {\cal{G}}_{ij} - \big(1-\lambda\big)g_{ij}\Delta\varphi \Big] = 8\pi G J_{i},
  \eq
where the super-momentum $\pi_{ij} $ and matter current $J_{i}$
are defined as
 \bqn \lb{eq2a}
\pi_{ij} &\equiv&   - K_{ij} + \lambda K g_{ij},\nb\\
J_{i} &\equiv& - N\frac{\delta{\cal{L}}_{M}}{\delta N^{i}}.
 \eqn
Similarly, variations of the action with respect to $\varphi$ and $A$ yield,
\bqn
\lb{eq4a}
& & {\cal{G}}^{ij} \Big(K_{ij} + \nabla_{i}\nabla_{j}\varphi\Big) + \big(1-\lambda\big)\Delta\Big(K + \Delta\varphi\Big) \nb\\
& & ~~~~~~~~~~~~~~~~~~~~~~~~~~~~~~~~ = 8\pi G J_{\varphi}, \\
\lb{eq4b}
& & R - 2\Lambda_{g} =   8\pi G J_{A},
\eqn
where
\bq
\lb{eq5}
J_{\varphi} \equiv - \frac{\delta{\cal{L}}_{M}}{\delta\varphi},\;\;\;
J_{A} \equiv 2 \frac{\delta\left(N{\cal{L}}_{M}\right)}{\delta{A}}.
\eq
On the other hand,  the dynamical equations now read,
 \bqn \lb{eq3}
&&
\frac{1}{N\sqrt{g}}\Bigg\{\sqrt{g}\Big[\pi^{ij} - \varphi {\cal{G}}^{ij} - \big(1-\lambda\big) g^{ij} \Delta\varphi\Big]\Bigg\}_{,t} 
\nb\\
& &~~~ = -2\left(K^{2}\right)^{ij}+2\lambda K K^{ij} \nb\\
& &  ~~~~~ + \frac{1}{N}\nabla_{k}\left[N^k \pi^{ij}-2\pi^{k(i}N^{j)}\right]\nb\\
& &  ~~~~~ - 2\big(1-\lambda\big) \Big[\big(K + \Delta\varphi\big)\nabla^{i}\nabla^{j}\varphi + K^{ij}\Delta\varphi\Big]\nb\\
& & ~~~~~ + \big(1-\lambda\big) \Big[2\nabla^{(i}F^{j)}_{\varphi} - g^{ij}\nabla_{k}F^{k}_{\varphi}\Big]\nb\\
& & ~~~~~ +  \frac{1}{2} \Big({\cal{L}}_{K} + {\cal{L}}_{\varphi} + {\cal{L}}_{A} + {\cal{L}}_{\lambda}\Big) g^{ij} \nb\\
& &  ~~~~~    + {\cal{F}}^{ij} + F_{\varphi}^{ij} +  F_{A}^{ij} + 8\pi G \tau^{ij},
 \eqn
where $\left(K^{2}\right)^{ij} \equiv K^{il}K_{l}^{j},\; f_{(ij)}
\equiv \left(f_{ij} + f_{ji}\right)/2$, and
 \bq \label{tau}
\tau^{ij} = {2\over \sqrt{g}}{\delta \left(\sqrt{g}
 {\cal{L}}_{M}\right)\over \delta{g}_{ij}}. 
 \eq
 The quantities $F^{i}_{\varphi},\; {\cal{F}}^{ij},\; F_{\varphi}^{ij}$ and $ F_{A}^{ij}$ are defined as, 
  \bqn
\lb{eq3a}
 {\cal{F}}^{ij} &\equiv&
\frac{1}{\sqrt{g}}\frac{\delta\left(-\sqrt{g}
{\cal{L}}_{V}\right)}{\delta{g}_{ij}}
 = \sum^{5}_{s=0}{\gamma_{s} \zeta^{n_{s}}
 \left({\cal{F}}_{s}\right)^{ij} },\nb\\
 F_{\varphi}^{i} &=&  \Big(K + \Delta\varphi\Big)\nabla^{i}\varphi + \frac{N^{i}}{N} \Delta\varphi, \nb\\
F_{\varphi}^{ij} &=&  \sum^{3}_{n=1}{F_{(\varphi, n)}^{ij}},\nb\\
F_{A}^{ij} &=& \frac{1}{N}\left[AR^{ij} - \Big(\nabla^{i}\nabla^{j} - g^{ij}\Delta\Big)A\right],
 \eqn
with 
$n_{s} =(2, 0, -2, -2, -2, -4)$ and 
  \bqn \lb{eq3b}
\left({\cal{F}}_{0}\right)_{ij} &=& \left({{F}}_{0}\right)_{ij},\;\;\;
\left({\cal{F}}_{1}\right)_{ij} = \left({{F}}_{1}\right)_{ij},\nb\\
\left({\cal{F}}_{2}\right)_{ij} &=&\left({{F}}_{2}\right)_{ij},\;\;\;
\left({\cal{F}}_{3}\right)_{ij} = \left({{F}}_{3}\right)_{ij}, \nb\\
 \left({\cal{F}}_{4}\right)_{ij} &=& - \epsilon^{kmn}\Bigg\{\nabla_m\nabla_{(i}\nabla_nR_{j)k}\nb\\
 & & +\nabla_l\Big[g_{(im}\nabla_{j)}\nabla_nR^l_k
  + g_{(in}R^l_mR_{j)k}\Big]\nb\\
& &  -\nabla^2\nabla_ng_{(im}R_{j)k}  - R_{m(i}\nabla_nR_{j)k}  \nb\\
& & +g_{ij}\nabla_l\nabla_k\nabla_nR^l_m\Bigg\},\\
 \left({\cal{F}}_{5}\right)_{ij} &=& \frac{1}{2} \left(F_{4}\right)_{ij}  -  \frac{5}{2} \left(F_{5}\right)_{ij}
 +3 \left(F_{6}\right)_{ij}\nb\\
 & & - \frac{3}{8} \left(F_{7}\right)_{ij} +  \left(F_{8}\right)_{ij},
 \eqn
where $\left(F_{s}\right)_{ij}$ are given in \cite{WM}. 

 The $F_{\varphi}^{ij}$'s are given by
 \bqn
   \lb{eq3c}
F_{(\varphi, 1)}^{ij} &=& \frac{1}{2}\varphi\Bigg\{\Big(2K + \Delta\varphi\Big) R^{ij}\nb\\
& & ~~~~~
- 2 \Big(2K^{j}_{k} + \nabla^{j} \nabla_{k}\varphi\Big) R^{ik} \nb\\
& & ~~~~~ - 2 \Big(2K^{i}_{k} + \nabla^{i} \nabla_{k}\varphi\Big) R^{jk}\nb\\
& &~~~~~
- \Big(2\Lambda_{g} - R\Big) \Big(2K^{ij} + \nabla^{i} \nabla^{j}\varphi\Big)\Bigg\},\nb\\
F_{(\varphi, 2)}^{ij} &=& \frac{1}{2}\nabla_{k}\left\{\varphi{\cal{G}}^{ik}
\Big(\frac{2N^{j}}{N} + \nabla^{j}\varphi\Big) \right. \nb\\
& &
+ \varphi{\cal{G}}^{jk}  \Big(\frac{2N^{i}}{N} + \nabla^{i}\varphi\Big)\nb\\
& &
-  \varphi{\cal{G}}^{ij}  \Big(\frac{2N^{k}}{N} + \nabla^{k}\varphi\Big)\Bigg\}, \nb\\
F_{(\varphi, 3)}^{ij} &=& \frac{1}{2}\Bigg\{2\nabla_{k} \nabla^{(i}f^{j) k}_{\varphi}
- \Delta f_{\varphi}^{ij} \nb\\
& & ~~~~~~~  - \left(\nabla_{k}\nabla_{l}f^{kl}_{\varphi}\right)g^{ij}\Bigg\},\nb\\
F_{(\varphi, 1)}^{ij} &=& \frac{1}{2}\varphi\Bigg\{\Big(2K + \Delta\varphi\Big) R^{ij}\nb\\
& & ~~~~~
- 2 \Big(2K^{j}_{k} + \nabla^{j} \nabla_{k}\varphi\Big) R^{ik} \nb\\
& & ~~~~~ - 2 \Big(2K^{i}_{k} + \nabla^{i} \nabla_{k}\varphi\Big) R^{jk}\nb\\
& &~~~~~
- \Big(2\Lambda_{g} - R\Big) \Big(2K^{ij} + \nabla^{i} \nabla^{j}\varphi\Big)\Bigg\},\nb\\
F_{(\varphi, 2)}^{ij} &=& \frac{1}{2}\nabla_{k}\left\{\varphi{\cal{G}}^{ik}
\Big(\frac{2N^{j}}{N} + \nabla^{j}\varphi\Big) \right. \nb\\
& &
+ \varphi{\cal{G}}^{jk}  \Big(\frac{2N^{i}}{N} + \nabla^{i}\varphi\Big)\nb\\
& &
-  \varphi{\cal{G}}^{ij}  \Big(\frac{2N^{k}}{N} + \nabla^{k}\varphi\Big)\Bigg\}, \nb\\
F_{(\varphi, 3)}^{ij} &=& \frac{1}{2}\Bigg\{2\nabla_{k} \nabla^{(i}f^{j) k}_{\varphi}
- \Delta f_{\varphi}^{ij} \nb\\
& & ~~~~~~~  - \left(\nabla_{k}\nabla_{l}f^{kl}_{\varphi}\right)g^{ij}\Bigg\},
\eqn
where
\bqn
\lb{eq3d}
f_{\varphi}^{ij} &=& \varphi\left\{\Big(2K^{ij} + \nabla^{i}\nabla^{j}\varphi\Big) 
- \frac{1}{2} \Big(2K + \Delta\varphi\Big)g^{ij}\right\}.\nb\\
\eqn

The matter, on the other hand, satisfies  the conservation laws,
 \bqn \lb{eq5a} & &
 \int d^{3}x \sqrt{g} { \left[ \dot{g}_{kl}\tau^{kl} -
 \frac{1}{\sqrt{g}}\left(\sqrt{g}J^{t}\right)_{, t}
 +   \frac{2N_{k}}  {N\sqrt{g}}\left(\sqrt{g}J^{k}\right)_{,t}
  \right.  }   \nb\\
 & &  ~~~~~~~~~~~~~~ \left.   - 2\dot{\varphi}J_{\varphi} -  \frac{A} {N\sqrt{g}}\left(\sqrt{g}J_{A}\right)_{,t}
 \right] = 0,\\
\lb{eq5b} & & \nabla^{k}\tau_{ik} -
\frac{1}{N\sqrt{g}}\left(\sqrt{g}J_{i}\right)_{,t}  - \frac{J^{k}}{N}\left(\nabla_{k}N_{i}
- \nabla_{i}N_{k}\right)   \nb\\
& & \;\;\;\;\;\;\;\;\;\;\;- \frac{N_{i}}{N}\nabla_{k}J^{k} + J_{\varphi} \nabla_{i}\varphi - \frac{J_{A}}{2N} \nabla_{i}A
 = 0.
\eqn

When the scalar field defined by Eqs.(\ref{5.11}) and (\ref{5.12}) is the only source,  we find that  
\bqn
\lb{5.13a}
J^t &=&  -2\Bigg(\frac{f}{2N^2}\Big(\dot{\chi}-N^k\nabla_k \chi\Big)^2+{\cal V}\Bigg) \nb\\
& & -\Big[c_1 \triangle \chi+c_2 \big(\nabla\chi\big)^{2}\Big]\big(\nabla\varphi\big)^{2}\nb\\
& & + f \Big[\big(\nabla^k\varphi\big)\big(\nabla_k\chi\big)\Big]^2,  \\
\lb{5.13b}
J^i &=& \frac{f}{N}\Bigg[\dot{\chi}-\Big(N^k + N \nabla^{k}\varphi\Big)\big(\nabla_k \chi\big)\Bigg] \nabla^i \chi \nb\\
      & &   + \Big[c_1 \triangle \chi+c_2 \big(\nabla\chi\big)^{2}\Big]\nabla^{i}\varphi,\\
\lb{5.13c} 
J_\varphi &=&\frac{1}{N\sqrt{g}}\Bigg\{\sqrt{g}\Big[c_1 \triangle \chi+c_2 \big(\nabla\chi\big)^{2}\Big]\Bigg\}_{,t}   \nb\\
&&-\frac{1}{N}\nabla_i\Bigg\{f\Big[\dot{\chi} - \big(N^{k} + N\nabla^{k}\varphi\big)\big(\nabla_{k}\chi\big)\Big]\nabla^{i}\chi\nb\\
& & ~~~~~~~~~ +  \Big[c_1 \triangle \chi+c_2 \big(\nabla\chi\big)^{2}\Big]\nb\\
& & ~~~~~~~~~~~ \times \Big(N^i+N\nabla^i\varphi\Big)\Bigg\}, \\
\lb{5.13d}
J_A &=& 2\Big[c_1 \triangle \chi+c_2 \big(\nabla\chi\big) ^{2}\Big],\\
\lb{5.13e}
\tau_{ij} &=& \tau_{ij}^{(0)} +\tau_{ij}^\varphi, 
\eqn
where  \cite{WWM}
\bqn
\lb{5.14}
 \tau^{(0)}_{ij}   &=&   g_{ij} \Big\{{\cal{L}}^{(0)}_{\chi} + \nabla_{k}\big[\big({\cal{V}}_{,1} +\Delta{\cal{V}}_{,2}\big)\nabla^{k}\chi  + {\cal{V}}_{,2}\nabla^{k}\Delta\chi\big]\Big\}\nb\\
 & & +  \big(1+ 2V_1+2V_5 {\cal P}_2\big)  \left(\nabla_{i} \chi\right) \left(\nabla_{j} \chi\right) \nb\\
& & - 2 \big(\nabla_{(i}{\cal{V}}_{,1}\big)\big(\nabla_{j)}\chi\big)- 2 \big(\nabla_{(i}\Delta{\cal{V}}_{,2}\big)\big(\nabla_{j)}\chi\big)\nb\\
& & - 2 \big(\nabla_{(i}{\cal{V}}_{,2}\big)\big(\nabla_{j)}\Delta\chi\big),\nb\\
\tau_{ij}^\varphi &=&  g_{ij}\Big\{{\cal{L}}^{(A,\varphi)}_{\chi} - \frac{1}{N}\nabla_{k}\big[c_{1}\big(A - {\cal{A}}\big)\nabla^{k}\chi\big]\Big\}\nb\\
& & + \frac{2({\cal{A}} - A)}{N}\Big[c_1\nabla_{i}\nabla_{j}\chi + c_2 \big(\nabla_{i}\chi\big)\big(\nabla_{j}\chi\big)\Big]\nb\\
& & + \Big[c_1\Delta\chi + c_2 \big(\nabla\chi\big)^{2}\Big]  \big(\nabla_{i}\varphi\big)\big(\nabla_{j}\varphi\big)\nb\\
& & + \frac{2f}{N}\Big[\dot{\chi} - \big(N^{k} + N\nabla^{k}\varphi\big)\big(\nabla_{k}\chi\big)\Big]\Big(\nabla_{(i}\chi\Big)\Big(\nabla_{j)}\varphi\Big)\nb\\
& & + \frac{2}{N}\nabla_{(i}\Big[c_{1}\big(A - {\cal{A}}\big)\nabla_{j)}\chi\Big],
 \eqn
and
 \bqn
 \lb{5.15} 
{\cal V}_{,1} &\equiv& \frac{\partial {\cal V}}{\partial
{\cal{P}}_{1}}
    =  2V_{2} {\cal{P}}_{1} + 3 V_{3} {\cal{P}}_{1}^{2}
    - \sigma^{2}_{3} {\cal{P}}_{2},\nb\\
{\cal V}_{,2} &\equiv& \frac{\partial {\cal V}}{\partial
{\cal{P}}_{2}}
    =   V_{4} + V_{5} (\nabla\chi)^2 - \sigma^{2}_{3} {\cal{P}}_{1}.
 \eqn

\section*{Appendix B: Cosmological Scalar Perturbations }

\renewcommand{\theequation}{B.\arabic{equation}} \setcounter{equation}{0}

In this Appendix, we first give a brief review of the flat FRW background, and then turn to consider scalar
perturbations. To have our results as much applicable as possible, in the following
 we do not restrict ourselves to any
specific matter fields. 

\subsection{Flat FRW Universe}

The  homogeneous and isotropic flat FRW universe in terms of the conformal time $\eta$ is given by Eq.(\ref{3.1}). 
The $U(1)$ gauge freedom of Eq.(\ref{2.3})  always allows us to set
$\hat{\varphi} = 0$.
Then, we find
\bqn
\lb{3.4}
\hat K_{ij} &=& - a{\cal{H}} \delta_{ij}, \;\;\; \hat R_{ij} = 0, \;\;\; \hat F^{i}_{\varphi} = 0,\nb\\
\hat F^{ij}_{A} &=&   \hat F^{ij}_{\varphi} = 0,\;\;\;
\hat F^{ij} =  - \frac{ \Lambda}{a^{2}} \delta^{ij},
\eqn
where ${\cal{H}} = {a'}/a, \;  \Lambda \equiv \zeta^{2} \gamma_{0}/2$, and $a' \equiv da/d\eta$.
Hence,  
 \bqn
\lb{3.6}
\hat{\cal{L}}_{K} &=&  3\big(1-3\lambda\big) \frac{{\cal{H}}^{2}}{a^{2}},\;\; \hat{\cal{L}}_{\varphi} = 0 = \hat{\cal{L}}_{\lambda}, \nb\\
\hat{\cal{L}}_{A} &=&  \frac{2\Lambda_{g}\hat{A}}{a}, \;\;\; \hat{\cal{L}}_{V} = 2\Lambda .
 \eqn
It can be shown that the super-momentum constraint (\ref{eq2}) is satisfied identically for
$\hat{J}^i = 0$, while the Hamiltonian constraint
(\ref{eq1}) yields,
 \bq \lb{3.7a}
\frac{1}{2}\big(3\lambda - 1\big)  \frac{{\cal{H}}^{2}}{a^{2}}   = \frac{8\pi G}{3} \hat\rho+ \frac{\Lambda}{3},
 \eq
where $\hat{J}^t \equiv -2\hat\rho$.
 On the other hand, Eqs.(\ref{eq4a}) and (\ref{eq4b}) give, respectively,
 \bqn
 \lb{3.8a}
 & & \Lambda_{g} {\cal{H}} = - \frac{8\pi G a}{3} \hat J_{\varphi},\\
 \lb{3.8b}
 & &   \Lambda_{g}=  - 4\pi G \hat J_{A},
 \eqn
while the dynamical equation (\ref{eq3}) reduces to 
 \bqn
 \lb{3.7b}
& &
 \frac{1}{2}\big(3\lambda - 1\big)\left(\frac{{a''}}{a^{3}}  - \frac{{\cal{H}}^{2}}{a^{2}}\right)=  - {4\pi G\over
3}(\hat\rho+3 \hat p) \nb\\
& & ~~~~~~~~~~~~~~~~~~~~~~~~~~~~~~~~~~ + {1\over3} \Lambda  - \frac{1}{2a}  \Lambda_{g}\hat A,
 \eqn
where   $\hat\tau_{ij} =  \hat p\,
\hat g_{ij}$.

The conservation law of momentum (\ref{eq5b}) is satisfied identically, while
the one of energy (\ref{eq5a}) reads,
 \bq \lb{3.8}
{{\hat\rho}'} + 3{\cal{H}} \left(\hat\rho + \hat p \right) = \hat A \hat J_{\varphi},
 \eq
which can be also obtained from Eqs.(\ref{3.7a}) and (\ref{3.7b}).

   \subsection{Linear  perturbations}

The linear scalar perturbations are given by Eq.(\ref{4.0a}). 
  Under the gauge transformations (\ref{1.4}), the perturbations transform as
\bqn
\lb{4.0b}
\tilde{\phi} &=& \phi - {\cal{H}}\xi^{0} - \xi^{0'},\;\;\;
\tilde{\psi} = \psi +  {\cal{H}}\xi^{0},\nb\\
\tilde{B} &=& B +  \xi^{0} - \xi',\;\;\;
\tilde{E} = E -   \xi,\nb\\
\tilde{\delta\varphi} &=& \delta\varphi - \xi^0 \hat{\varphi}',\;\;\;
\tilde{\delta{A}} = \delta{A} - \xi^0 \hat{A}' - \xi^{0'} \hat{A}, ~~~
\eqn
where $f = - \xi^0,\; \zeta^i = - \xi^{,i}, \;  {\cal{H}} \equiv a'/a$, and a prime denotes the ordinary derivative with respect
to $\eta$. Under the $U(1)$ gauge transformations, on the other hand,
we find that
\bqn
\lb{4.0c}
\tilde{\tilde{\phi}} &=& \tilde{\phi},\;\;\; 
\tilde{\tilde{E}} = \tilde{E},\;\;\;
\tilde{\tilde{\psi}} = \tilde{\psi},\;\;\;
\tilde{\tilde{B}} =  \tilde{B} - \frac{\epsilon}{a}, \nb\\
\tilde{\tilde{\delta\varphi}} &=& \tilde{\delta\varphi} + \epsilon,\;\;\;
\tilde{\tilde{\delta{A}}} = \tilde{\delta{A}} - \epsilon',
%
\eqn
where $\epsilon = - \alpha$. Then, the  gauge transformations of the whole group $ U(1) \ltimes {\mbox{Diff}}(M, \; {\cal{F}})$
will be the linear combination of the above two. Out of the six unknown, one can construct three
gauge-invariant quantities,
\bqn
\lb{4.0d}
\Phi &=& \phi - \frac{1}{a - \hat{\varphi}'}\big(a\sigma - \delta\varphi\big)' \nb\\
& & -  \frac{1}{\big(a - \hat{\varphi}'\big)^2}\big(\hat{\varphi}'' - {\cal{H}}\hat{\varphi}'\big)
\big(a\sigma- \delta\varphi\big),\nb\\
\Psi &=& \psi +  \frac{{\cal{H}}}{a - \hat{\varphi}'}\big(a\sigma- \delta\varphi\big),\nb\\
\Gamma &=& \delta{A} +\Bigg[\frac{a\big(\delta\varphi - \hat{\varphi}'\sigma\big) - \hat{A}\big(a\sigma - \delta\varphi\big)}{a - \hat{\varphi}'}\Bigg]',
\eqn
where  $\sigma\equiv E'-B$. For the background, we have chosen the gauge (\ref{gauge}), for
which Eq.(\ref{4.0d}) reduces to
\bqn
\lb{4.0db}
\Phi &=& \phi  - \frac{1}{a}\big(a\sigma - \delta\varphi\big)' , \nb\\
\Psi &=& \psi - \frac{{\cal{H}}}{a}\big(\delta\varphi -a\sigma\big),\nb\\
\Gamma &=& \delta{A} + \Bigg[\delta\varphi - \frac{\hat{A}}{a}\big(a\sigma -  \delta\varphi\big)\Bigg]',\; (\hat{\varphi} = 0).
\eqn


 Then, with the gauge choice,
\bq
\lb{4.0ddd}
{\delta\varphi} = \phi  = E = 0, 
\eq
 to first-order  it can be shown that the Hamiltonian and momentum constraints become,  respectively,
 \bqn \lb{4.4}
&&  \int d^{3}x\Big\{\partial^2\psi - \frac{1}{2}(3\lambda-1){\cal{H}}(3{\psi'} +\partial^2B)\nb\\
&& ~~~~~~~~~~~~ - 4\pi G a^{2}\delta\mu\Big\}=0,\\
 \lb{4.5}
&&(3\lambda-1){\psi'} + (\lambda - 1)\partial^2B=8\pi G a q, ~~~~
 \eqn
where 
\bq
\lb{4.5a}
\delta\mu \equiv -\frac{1}{2}\delta{J^{t}},\;\;\; 
 \delta{J}^{i} \equiv  \frac{1}{a^{2}}q^{,i}.
 \eq

On the other hand, the linearized equations (\ref{eq4a}) and (\ref{eq4b}) reduce, respectively,  to
 \bqn
 \lb{4.6a}
&& 2{\cal{H}} \partial^{2}\psi + \Big[\Lambda_{g}a^{2} + \big(1-\lambda\big)\partial^{2}\Big]\Big(3\psi' + \partial^{2}B\Big)\nb\\
&&~~~~~~~~~~~~~~~~~~  = 8\pi G a^{3} \delta J_{\varphi},\\ 
 \lb{4.6b}
&&  \partial^2\psi = 2\pi G a^{2} \delta J_{A}.
 \eqn

 The linearized  dynamical equations can be divided into the trace and traceless parts. The trace part  reads,
 \bqn
\lb{4.7a}
&& \psi'' + 2{\cal{H}}\psi' + \frac{1}{3}\partial^{2}\Big(B' + 2{\cal{H}}B\Big) = \frac{1}{3\lambda -1}\Bigg\{8\pi G a^{2}\delta{\cal{P}}\nb\\
&& ~~~~+ \frac{2}{3}\Bigg[\frac{a -  \hat{A}}{a} + \frac{8\gamma_{2} + 3\gamma_{3}}{\zeta^{2} a^{2}}\partial^{2}\Bigg]\partial^{2}\psi\nb\\
&& ~~~~ + \frac{1}{a} \Bigg(\Lambda_{g} a^{2} + \frac{2}{3}\partial^{2}\Bigg)\delta{A}\Bigg\},
\eqn
 while the traceless part   is given by
 \bqn
 \label{4.7b}
&& B' + 2{\cal{H}}B - \Bigg[\frac{a -  \hat{A}}{a}+ \frac{8\gamma_{2} + 3\gamma_{3}}{\zeta^{2} a^{2}}\partial^{2}\Bigg]\psi\nb\\
& & ~~~~~~~~~~~~~~ - \frac{1}{a}\delta{A} = - 8\pi G a^{2}\Pi,
\eqn
where
\bqn
\lb{4.7aa}
\delta{\tau}^{ij} &=& \frac{1}{a^{2}}\Big[\big(\delta{\cal{P}} + 2\hat{p}\psi\big)\delta^{ij} + \Pi^{,<ij>}\Big],\nb\\
\Pi^{,<ij>} &=& \Pi^{, ij} - \frac{1}{3}\delta^{ij} \partial^{2}\Pi.  
\eqn

 The conservation laws (\ref{eq5a}) and (\ref{eq5b}) to first order are  given by,
 \bqn \lb{4.9a}
& & \int  d^{3}x \Bigg\{\delta\mu' + 3{\cal{H}} \delta\mu - 3(\hat{\rho} + \hat{p})\psi' + 3 {\cal{H}}\delta{\cal{P}}\nb\\
& & + \frac{1}{2a}\Big[3\hat{A}\hat{J}_{A}\psi' - \hat{A} \big(\delta{J}_{A}' + 3{\cal{H}} \delta{J}_{A}\big)\nb\\
& & ~~~~~~~~ -  2 a \hat{J}_{\varphi} \delta{A}\Big]\Bigg\} = 0,\\
\lb{4.9b}
& & q'  + 3{\cal{H}} q + \frac{1}{2}\hat{J}_{A}\delta{A} = a\Bigg(\delta{\cal{P}} + \frac{2}{3}\partial^{2}\Pi\Bigg),
 \eqn
where $\hat{J}_{A}$  is given by Eq.(\ref{3.8b}). 

This completes the general description of linear scalar perturbations in the flat FRW background
 in the framework of the HMT setup with detailed balance condition softly breaking and any given $\lambda$, 
 generalized recently by da Silva \cite{Silva}.

\section*{Appendix C: Linear Perturbations for $\lambda = 1$ and $c_{1} \not= 0$ }

\renewcommand{\theequation}{C.\arabic{equation}} \setcounter{equation}{0}

When  $\lambda = 1$ and $c_{1} \not= 0$,  Eqs.(\ref{d.1}) and (\ref{d.3}) yields,
\bq
\lb{c.14}
c_{1}\delta\chi' + \big(c_{1}' - f \hat{\chi}'\big)\delta\chi = 0,
\eq
which has the solution,
\bq
\lb{c.15}
\delta\chi = \frac{1}{c_{1}}\exp\Big[\delta\chi_{0}(\eta) + \delta\chi_{1}(x)\Big],
\eq
where $\delta\chi_{0}(\eta) = f\int{(\hat{\chi}'/c_{1})d\eta}$, and $ \delta\chi_{1}(x)$ is an arbitrary function of $x^{i}$ only. 
Inserting the above solution into Eqs.(\ref{d.3}), (\ref{d.4}) and (\ref{d.5}), one finds,
\bqn
\lb{c.16}
\psi &=&    {4\pi G}\exp\Big[\delta\chi_{0}(\eta) + \delta\chi_{1}(x)\Big],\nb\\
B_{k} &=& \frac{e^{-\delta\chi_{0}}}{a^{2}}\Bigg[\int{Qe^{\delta\chi_{0}}d\eta } + {\cal{R}} \left(x^{i}\right)\Bigg],\nb\\
\delta{A}_{k} &=& - \frac{a}{c_{1}}\left(f\hat{\chi}'B_{k} + P \delta\chi\right),
\eqn
where ${\cal{R}}(x)$ is another arbitrary function of $x^{i}$ only, and 
\bqn
P &\equiv& \frac{f}{k^{2}}\left[\frac{\hat{\chi}'^{2}}{c^{2}_{1}}\left(2c_{1}'^{2} + f^{2} - 3fc_{1}' - c_{1}c_{1}''\right) \right.\nb\\
& & \left. ~~~~~ + \frac{\hat{\chi}'' + 2{\cal{H}}\hat{\chi}' }{c_{1}}\left(f - c_{1}'\right)  - 12\pi G f \hat{\chi}'^{2}\right]\nb\\
&& ~~~~~ + \frac{1}{a}\left[a(1 + 2V_{1}) + 2\hat{A}\left(c_{1}' - c_{2}\right)\right]\nb\\
&& ~~~~~ + \frac{a^{2}}{k^{2}}V''  + \frac{2k^{2}}{a^{2}}\left(V_{2} + V_{4}'\right)  + \frac{2\sigma^{2}_{3}k^{4}}{a^{4}},\nb\\ 
Q &\equiv& \frac{\delta\chi}{c_{1}}\left[ 4\pi G c^{2}_{1}\left( a^{2} - a\hat{A} - \frac{8\gamma_{2} + 3\gamma_{3}}{\zeta^{2}}k^{2}\right)\right.\nb\\
& & ~~~~~~~ - a^{2} P\Bigg].
 \eqn





\begin{thebibliography}{nbound}


\bibitem{Horava}
P. Horava, JHEP, {\bf 0903}, 020 (2009) [arXiv:0812.4287]; Phys.
Rev. D{\bf 79}, 084008 (2009) [arXiv:0901.3775]; and Phys. Rev.
Lett. {\bf 102}, 161301  (2009) [arXiv:0902.3657].

\bibitem{Visser} M. Visser, Phys. Rev. D{\bf 80}, 025011 (2009) [arXiv:0902.0590];   arXiv:0912.4757.

\bibitem{ADM}  C.W. Misner, K.S. Thorne, and J.A. Wheeler, {\em Gravitation } (W.H. Freeman and Company, San Francisco, 1973),
pp.484-528. 

 \bibitem{LP}        M. Li and Y. Pang,   J. High Energy Phys. {\bf 08}, 015 (2009) [arXiv:0905.2751];
                             M. Henneaux, A. Kleinschmidt, and G.L. G—mez,  Phys. Rev. D{\bf 81}, 064002 (2010) [arXiv:0912.0399].

\bibitem{Cardy}  J. Cardy, Scaling and Renormalization in Statistical Physics (Cambridge University Press, Cambridge, 1966).

 
 \bibitem{HWW} Y.-Q. Huang, A. Wang, and Q. Wu,  Mod. Phys. Lett. {\bf 25}, 2267  (2010) [arXiv:1003.2003].

\bibitem{Mukc} S. Mukohyama,  Class. Quantum Grav. {\bf 27}, 223101 (2010) [arXiv:1007.5199].

 \bibitem{Sotiriou}  T.P. Sotiriou, arXiv:1010.3218.


  \bibitem{Padilla}   A. Padilla,   J. Phys. Conf. Ser. {\bf 259}, 012033 (2010) [arXiv:1009.4074].
  
   \bibitem{Hreview}        P. Horava, arXiv:1101.1081.        
 
                      
  
  \bibitem{SVW}  T. Sotiriou, M. Visser, and S. Weinfurtner,  Phys. Rev. Lett. {\bf 102},  251601 (2009)
                        [arXiv:0904.4464]; J. High Energy Phys.,  {\bf 10}, 033 (2009) [arXiv:0905.2798];  arXiv:1002.0308.


 \bibitem{WM} A. Wang and R. Maartens,  Phys. Rev. D {\bf 81}, 024009 (2010) [arXiv:0907.1748].  
 
 \bibitem{Wanga} A. Wang, Phys. Rev. D{\bf 82}, 124063 (2010) [arXiv:1008.3637].

 \bibitem{HW} Y.-Q. Huang and A. Wang, Phys. Rev. D,  {\em in press}  (2011) [arXiv:1011.0739].          
 

\bibitem{OS} R. Iengo, J.G. Russo, and M. Serone, JHEP, {\bf 11}, 020 (2009) [arXiv:0906.3477]; 
D. Orlando and S. Reffert, Class. Quantum Grav. {\bf 26}, 155021 (2009) [arXiv:0905.0301]. 


\bibitem{SC}      C. Charmousis, G. Niz, A. Padilla, and P.M. Saffin, J. High Energy Phys.,  {\bf 08}, 070 (2009) [arXiv:0905.2579].
                       
\bibitem{BPS}         D. Blas, O. Pujolas, and S. Sibiryakov, J. High Energy Phys.,  {\bf 03}, 061 (2009) [arXiv:0906.3046]. 

\bibitem{KA}   K. Koyama and F. Arroja, J. High Energy Phys., {\bf 03}, 061 (2010) [arXiv:0910.1998].
                       
 \bibitem{PS}      A. Papazoglou and T.P. Sotiriou, Phys. Lett. B{\bf 685}, 197 (2010) [arXiv:0911.1299]. 

 \bibitem{KP}      I. Kimpton and A. Padilla, J. High Energy Phys. {\bf 07}, 014 (2010) [arXiv:1003.5666].
                      

\bibitem{WWb} A. Wang and Q. Wu, Phys. Rev. D{\bf 83}, 044025 (2011) [arXiv:1009.0268].




  \bibitem{BPSc}   	 D.Blas, O.Pujolas, and S. Sibiryakov,  Phys. Rev. Lett. {\bf 104}, 181302 (2010) [arXiv:0909.3525];
                          JHEP, {\bf 1104}, 018 (2011) [arXiv.1007.3503].
                         
                         
                      
  \bibitem{Vain}  A. I. Vainshtein, Phys. Lett.  B{\bf 39}, 393  (1972).
  
                 
                      
 \bibitem{BS}     C. Bogdanos, and E.N. Saridakis,  Class. Quantum Grav. {\bf 27}, 075005 (2010) [ arXiv:0907.1636].        
 
  \bibitem{Wang}  A. Wang, Mod. Phys. A{\bf 26}, 387 (2011) [arXiv:1003.5152].             
                      
                      
 \bibitem{CalA} G. Calcagni,  J. High Energy Phys., {\bf 09}, 112 (2009) [arXiv:0904.0829].                       
                      
 \bibitem{LMP}                 H. L\"u, J. Mei, and C.N. Pope,  Phys. Rev. Lett. {\bf 103}, 091301 (2009) [arXiv:0904.1595].
                      
                      
                      
    \bibitem{AdSCFT}   J.M. Maldacena,  Adv. Theor. Math. Phys.  {\bf 2},  231 (1998); O. Aharony, S.S. Gubser, J. Maldacena,  
    H. Ooguri, and Y. Oz,  Phys. Rept.  {\bf 323}, 183 (2000).
    
    
  \bibitem{OM} L. Onsager and S. Machlup, Phys. Rev. {\bf 91}, 1505 (1953); S. Machlup and L. Onsager, {\em ibid.}, {\bf 91}, 1512 (1953). 
    
         
      \bibitem{Verlinde}  E.P. Verlinde, arXiv:1001.0785.
    
    
   \bibitem{HMT} P. Horava and C.M. Melby-Thompson, Phys. Rev. D{\bf 82}, 064027 (2010) [arXiv:1007.2410].
   
   
    \bibitem{WW}  A. Wang and Y. Wu,   Phys. Rev. D{\bf 83}, 044031 (2011) [arXiv:1009. 2089].
    
 \bibitem{Silva} A.M. da Silva, Class. Quantum Grav. {\bf 28}, 055011 (2011) [arXiv:1009.4885].
 
 
  \bibitem{Kluson}     J. Kluson, Phys. Rev. D{\bf 83}, 044049 (2011) [arXiv:1011.1857];   arXiv:1101.5880; 
J. Kluson, S. Nojiri, S.D. Odintsov, and D. Saez-Gomez,  arXiv:1012.0473. 

                      
\bibitem{KK}   A. Kehagias and K. Sfetsos, Phys. Lett. B{\bf 678}, 123 (2009) [arXiv:0905.0477]; 
                  M.-i. Park,  J. High Energy Phys. {\bf 09},   123 (2009) [arXiv:0905.4480];   G.Koutsoumbas, E. Papantonopoulos, P. Pasipoularides,
                  and M.Tsoukalas,  Phys.Rev. D{\bf 81}, 124014 (2010) [arxiv:1004.2289]; and references therein. 

\bibitem{Cai}   R. G.	Cai,	L. M.	 Cao, and	N. Ohta, Phys. Rev. D{\bf 80}, 024003 (2009); Phys. Lett. B {\bf 679}, 504 (2009); 
R. G.	Cai and N. Ohta, Phys. Rev.  D {\bf 81}, 084061 (2010); and references therein. 

 \bibitem{AP} J.  Alexandre and P. Pasipoularides,   Phys. Rev. D{\bf 83}, 084030 (2011) [arXiv:1010.3634]. 
  
  \bibitem{GSW} J. J. Greenwald, V.H. Satheeshkumar, and A. Wang, JCAP, {\bf 12}, 007 (2010) [arXiv:1010.3794].


 \bibitem{WWM} A. Wang, D. Wands, and R. Maartens,      J. Cosmol. Astropart. Phys.,  {\bf 03}, 013  (2010) [arXiv:0909.5167].
 

\bibitem{KKb} E. Kiritsis and G. Kofinas,  Nucl. Phys. B{\bf 821}, 467 (2009) [arXiv:0904.1334].


   \bibitem{TS} T. Takahashi and J. Soda, Phys. Rev. Lett. {\bf 102}, 231301 (2009) [arXiv:0904.0554 [hep-th].



 \bibitem{CalB} G. Calcagni,     Phys. Rev. D{\bf 81}, 044006 (2010) [arXiv:0905.3740].
 
 
\bibitem{MW09} K.A. Malik and D. Wands, Phys. Reports {\bf 475}, 1 (2009).

  
 %
            

 
 



\end{thebibliography}
\end{document}